\documentclass[twocolumn]{aastex631}

\usepackage{url}

\usepackage{amsmath}
\usepackage{float}
\usepackage{rotating}
\usepackage{hyperref}

\newcommand{\rstar}{\ensuremath{R_{*}}}

\graphicspath{{./}{Figures/}}

\shorttitle{SOLES VII: The Spin-Orbit Alignment of WASP-106 b}
\shortauthors{Wright et al.}

\begin{document}
\title{SOLES VII: The Spin-Orbit Alignment of WASP-106 b, a Warm Jupiter Along the Kraft Break}

\author[0000-0003-1422-2977]{Josette Wright}
\affiliation{Department of Astronomy, Indiana University, Bloomington, IN 47405, USA}

\author[0000-0002-7670-670X]{Malena Rice}
\affiliation{Department of Astronomy, Yale University, New Haven, CT 06511, USA}

\author[0000-0002-0376-6365]{Xian-Yu Wang}
\affiliation{Department of Astronomy, Indiana University, Bloomington, IN 47405, USA}

\author[0000-0002-8685-5397]{Kyle Hixenbaugh}
\affiliation{Department of Astronomy, Indiana University, Bloomington, IN 47405, USA}

\author[0000-0002-7846-6981]{Songhu Wang}
\affiliation{Department of Astronomy, Indiana University, Bloomington, IN 47405, USA}

\correspondingauthor{Josette Wright}
\email{jowrig@iu.edu}

\begin{abstract}
Although close-orbiting, massive exoplanets -- known as hot and warm Jupiters -- are among the most observationally accessible known planets, their formation pathways are still not universally agreed upon. One method to constrain the possible dynamical histories of such planets is to measure the systems' sky-projected spin-orbit angles using the Rossiter-McLaughlin effect. By demonstrating whether planets orbit around the stellar equator or on offset orbits, Rossiter-McLaughlin observations offer clues as to whether the planet had a quiescent or violent formation history. Such measurements are, however, only a reliable window into the history of the system if the planet in question orbits sufficiently far from its host star; otherwise, tidal interactions with the host star can erase evidence of past dynamical upheavals. We present a WIYN/NEID Rossiter-McLaughlin measurement of the tidally detached ($a/R_* = 13.18^{+0.35}_{-0.37}$) warm Jupiter WASP-106 b, which orbits a star along the Kraft break ($T_{\mathrm{eff}}=6002\pm164$ K). We find that WASP-106 b is consistent with a low spin-orbit angle ($\lambda=6^{+17}_{-16}\,\degr$ and $\psi = 26^{+12}_{-17}\,\degr$), suggesting a relatively quiescent formation history for the system. We conclude by comparing the stellar obliquities of hot and warm Jupiter systems, with the WASP-106 system included, to gain insight into the possible formation routes of these populations of exoplanets.
\end{abstract}


\section{Introduction} 
\label{section:intro}

Despite being among the most readily detectable exoplanets, short-period Jovian planets still have contested formation histories. The formation pathways available to this population have been the subject of much debate in the last several decades, as reviewed in \citet{dawson2018origins}. One major category of formation routes is violent formation, in which the Jupiter originally forms farther out from the star than its final, close-in orbit and then moves inward through high-eccentricity migration, causing disruptions to the orbits of inner planets on its way \citep{Weidenschilling1996_grav_scattering, wu2003planet, fabrycky2007shrinking, Ida2013_eccentricity}. Quiescent formation pathways, on the other hand, may better preserve nearby companions in hot Jupiter systems \citep{lee2002dynamics}. In situ formation, for example, can occur when a less massive super-Earth in the inner region of a planetary system undergoes runaway gas accretion until it becomes a Jupiter \citep{batygin2016situ}. Another possible quiescent formation route involves disk migration of the planet while already at its Jupiter-range mass \citep{Goldreich_1979_diskwaves, Lin1996_migration, Baruteau2014_migration}. Information about the current properties of known hot Jupiter systems may be used to distinguish between these possible origins.

One parameter that offers a window into the dynamical histories of these systems is the sky-projected spin-orbit alignment angle, $\lambda$, which is a proxy for the stellar obliquity. The stellar obliquity ($\psi$) of a planetary system is defined as the angle between the net orbital angular momentum axis of the planetary system and the spin axis of its host star. A large angle $\lambda$ typically corresponds to a misalignment between these two vectors and may indicate a violent past. Conversely, an aligned system with a smaller $\lambda$ would be consistent with a more quiescent history.

To measure $\lambda$, we utilize the Rossiter-McLaughlin effect \citep{rossiter1924detection, mclaughlin1924some}, which describes the way in which a body transiting in front of a rotating star blocks out the blue- and red-shifted portions of the occulted star’s light at different points during the transit. The proportions of blue- and red- shifted light that are blocked can be traced through high-precision radial velocity (RV) observations, and the shape of the observed RV profile encodes information about the sky-projected spin-orbit angle $\lambda$ at which the transiting body crosses in front of the star. 

The majority of spin-orbit angle measurements to date have been made for hot Jupiter systems due to the planets' relatively deep and frequent transits, which makes them observationally accessible \citep{albrecht2022stellar}. However, because hot Jupiters orbit so close to their host star, tidal effects may, in some cases, erase the remnant signatures of a chaotic dynamical past \citep{winn2010hot}. Previous results have demonstrated that hot Jupiters orbiting hot stars above the Kraft break \citep{kraft1967studies} -- a rotational discontinuity that divides stars with convective envelopes (cool stars) and those with radiative envelopes (hot stars) -- are misaligned significantly more often than hot Jupiters orbiting cooler stars, below the Kraft break \citep{winn2010hot, schlaufman2010evidence}. The Kraft break is located in the range $6000 \leq T_{\rm eff} \leq 6250\, \rm K$, with some stellar-metallicity-dependent variation in the exact transition point \citep{Spalding_2022}. 

Because cooler stars have a substantial convective envelope, a popular explanation for the observed trend in stellar obliquities is that hot Jupiters orbiting stars below the Kraft break are tidally realigned, while their counterparts orbiting hotter stars remain misaligned due to their weaker star-planet tidal interactions \citep[e.g.][]{albrecht2012obliquities, wang2021aligned}. This scenario suggests that hot Jupiters may form violently in both hot and cool star systems, producing regular spin-orbit misalignments across stellar types. In this framework, the signatures of misalignment are rapidly erased from cool star systems, while they often persist for hot star systems.

Previous work has demonstrated that this violent formation mechanism, combined with tidal dissipation, can well reproduce the observed set of hot Jupiter spin-orbit angles \citep{rice2022origins}. However, it remains unclear whether hot Jupiters around cool stars did indeed begin with larger initial misalignments. This leads to major degeneracies when determining the system’s history, as reviewed in Section 4 of \citet{albrecht2022stellar}. 

One way to break this degeneracy is to obtain spin-orbit measurements for wider-orbiting, ``tidally detached'' systems -- that is, those that have projected tidal realignment timescales longer than the age of the system \citep{Rice2021SOLESI}. This is the goal of the Stellar Obliquities in Long-period Exoplanet Systems (SOLES) survey \citep{Rice2021SOLESI, wang2022aligned, rice2022tendency, rice2023orbital, hixenbaugh2023spin, dong2023toi}, which is collecting stellar obliquity measurements for systems with tidally detached planets to more robustly constrain the origins of spin-orbit misalignments in exoplanet systems. By expanding the set of spin-orbit constraints for tidally detached warm Jupiters -- which we define as short-period ($P<100$ days) giant planets ($M_p \geq 0.4 M_J$) with scaled orbital semimajor axes $a/R_* \geq 11$ -- we can place new constraints on the origins of spin-orbit misalignments more generally.


In this work, we present a measurement of the Rossiter-McLaughlin effect across one transit of the tidally detached ($a/R_* = 13.18^{+0.35}_{-0.37}$) warm Jupiter WASP-106 b, first confirmed by \citet{smith2014wasp}. This observation was taken with the NEID spectrograph \citep{Schwab2016NEID} mounted on the WIYN 3.5-meter telescope at Kitt Peak National Observatory in Arizona. This is the seventh result from the ongoing SOLES survey and one of the first warm Jupiter spin-orbit angles measured in a system with a relatively hot, high-mass host star along the Kraft break. WASP-106 b is a $1.93\pm 0.15 \, \rm{M_J}$ planet, orbiting a $M_*=1.175^{+0.082}_{-0.074} \,\mathrm{M_{\odot}}$, $T_{\mathrm{eff}}=6002\pm164$ K star at a period of $P=9.29$ days (see Section \ref{section:spinorbitmodel}). We find that WASP-106 b is consistent with near-alignment, with $\lambda=6^{+17}_{-16}\,\degr$ and $\psi=26^{+12}_{-17}\degr$.

\section{Observations}
\label{section:observations}
On March 2nd, 2022, from 4:34 to 11:57 UT, we collected 22 RV measurements of WASP-106 using the high-resolution ($R\sim110,000$) WIYN/NEID spectrograph which covers a wavelength range 380-930 nm \citep{Schwab2016NEID}. Seeing ranged from 1.0$\arcsec$ to 1.7$\arcsec$, with a median of 1.1$\arcsec$, and the median RV uncertainty was 9.3 m/s. The airmass $z$ varied within the range $1.26 \leq z \leq 1.78$, starting at $z=1.58$ at the beginning of the night before the target rose to $z=1.26$ and then reached $z=1.78$ at the end of the observing sequence. The signal-to-noise ratio (S/N) ranged from 18 to 30 pixel$^{-1}$ at 5500 \AA.

The spectra from this observing sequence were reduced using the NEID Data Reduction Pipeline\footnote{See https://neid.ipac.caltech.edu/docs/NEID-DRP/ for more information}, and reduced spectra were retrieved from the NExScI NEID Archive\footnote{https://neid.ipac.caltech.edu/}.
The NEID RV measurements and uncertainties are provided in Table \ref{tab:neid_data} and are shown in the rightmost panel of Figure \ref{fig:rv_joint_fit}.

\begin{deluxetable}{lll}
\tablecaption{NEID radial velocity measurements across the transit of WASP-106 b. For readability, values are reported as RV - 17200 m/s.
\label{tab:neid_data}}
\tabletypesize{\scriptsize}
\tablehead{\colhead{Time ({BJD$_{\rm TDB}$})} & \colhead{RV - 17200 (m/s)} & \colhead{$\sigma_{\rm RV}$ (m/s)}}
\startdata
2459640.6907652 & 1.0 & 12.2\\
2459640.7048980 & 7.4 & 9.6\\
2459640.7182308 & 14.1 & 10\\
2459640.7324765 & 4.8 & 9.6\\
2459640.7460450 & 22.7 & 11.3\\
2459640.7614610 & 9.6 & 9.5\\
2459640.7749805 & 19.0 & 9.1\\
2459640.7895711 & 18.6 & 9.4\\
2459640.8043050 & 13.3 & 8.4\\
2459640.8284706 & -3.0 & 9.0\\
2459640.8420965 & -20.6 & 9.3\\
2459640.8565140 & -25.2 & 8.8\\
2459640.8713091 & -33.7 & 8.5\\
2459640.8851927 & -47.0 & 8.0\\
2459640.8992462 & -36.5 & 8.2\\
2459640.9137848 & -36.2 & 8.8\\
2459640.9276972 & -53.1 & 9.0\\
2459640.9415950 & -32.5 & 7.8\\
2459640.9571293 & -46.2 & 10.3\\
2459640.9701108 & -28.4 & 10.7\\
2459640.9844582 & -4.1 & 10.6\\
2459640.9979400 & -2.4 & 12.1\\
\enddata
\end{deluxetable}

\section{Stellar Parameters} 
\label{section:stellar_parameters}  

We derived the stellar parameters for WASP-106 by analyzing the NEID spectra obtained during the RM sequence. To enhance the final spectrum S/N, we co-added all spectra after correcting for their RV shifts caused by the planetary reflex motion. The stellar parameters $T_{\rm eff}$, $\rm{log}\,g$, $\rm {[Fe/H]}$, and $v\sin{i_*}$ were derived using the \texttt{iSpec} Python package \citep{blanco2014ispec, blanco2019ispec}.

During the fitting process, we employed the SPECTRUM radiative transfer code \citep{Gray1994}, MARCS atmosphere models \citep{gustafsson2008_MARCS}, and the sixth version of the GES atomic line list \citep{Heiter2021_GES}. Using constraints from these sources, \texttt{iSpec} minimizes the difference between the synthetic and input spectra by applying the nonlinear least-squares Levenberg-Marquardt fitting algorithm \citep{more2006levenberg}.

To expedite the fitting process, we selected specific spectral regions from 476 to 678 nm that are sensitive to our parameters of interest. These regions include the wing segments of the H$\alpha$, H$\beta$, and Mg I triplet lines, which are sensitive to $T_{\mathrm{eff}}$ and $\log g$. Additionally, we included the Fe I and Fe II lines, which enable precise constraints on [Fe/H] and $v\sin i_*$.

To determine the stellar mass ($M_{*}$) and radius ($R_{*}$), we employed MESA Isochrones \& Stellar Tracks \citep[MIST;][]{Choi2016mist,Dotter2016mist} models in conjunction with a spectral energy distribution (SED) fit using the \texttt{EXOFASTv2} Python package \citep{eastman2019exofast}. The SED was constructed using photometry from 2MASS \citep{Cutri2003}, WISE \citep{Cutri2013}, TESS \citep{ricker2015tess}, and \textit{Gaia} DR3 \citep{GaiaCollaboration2023}.

During the fitting process, we adopted Gaussian priors based on the effective temperature ($T_{\rm eff}$) and metallicity ($\rm {[Fe/H]}$) derived from our spectral fit, as well as the parallax ($\varpi$) drawn from \textit{Gaia} DR3 \citep{GaiaCollaboration2023}. We also applied an upper limit on the V-band extinction ($A_v$) as given by \citet{Schlafly2011}.

The fitting process utilized the Differential Evolution Markov Chain Monte Carlo \citep[DEMCMC;][]{ter2006markov} method. The fit was considered converged when the Gelman-Rubin statistic ($\hat{R}$, \citealt{Gelman1992}) fell below 1.01 and the effective number of independent samples exceeded 1000. The resulting stellar parameters are listed in Table~\ref{tab:system_properties}.\footnote{This method provided an intermediate value of $v\sin i_*=5.83\pm3.61$ km/s that was used to inform our priors to \texttt{allesfitter}, from which we derived the final $v\sin i_*$ value in Table~\ref{tab:system_properties}.}

\section{Stellar Obliquity Modeling} 
\label{section:spinorbitmodel}

\begin{figure*}
    \centering
    \includegraphics[width=1.0\linewidth]{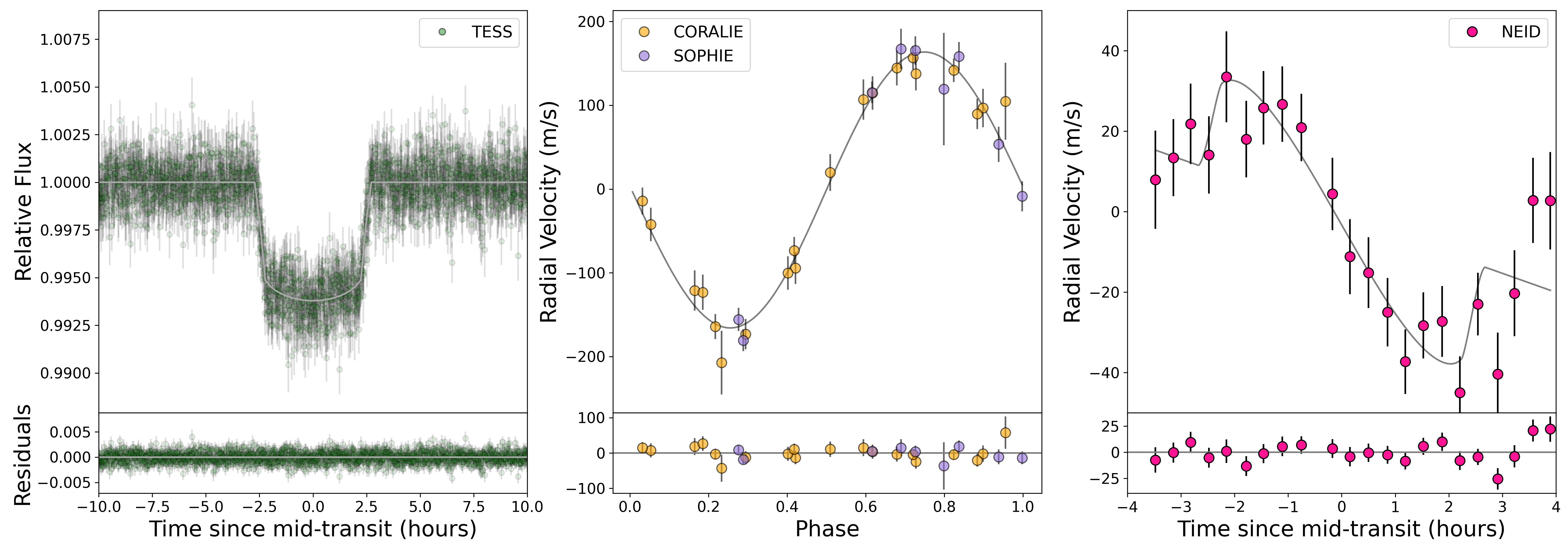}
    \caption{Joint fit to photometry (left), out-of-transit RV data (center), and the in-transit Rossiter-McLaughlin RV data (right) obtained for WASP-106 b. The model is shown in gray, while data is provided in color with modeled constant offsets and jitter terms included. The associated residuals are provided below each panel.}
    \label{fig:rv_joint_fit}
\end{figure*}

To find the sky-projected spin-orbit angle $\lambda$ for WASP-106 b, we used the Python package \texttt{allesfitter} \citep{gunther2020allesfitter} to jointly fit radial velocity data from the NEID \citep{Schwab2016NEID}, CORALIE \citep{queloz2000coralie}, and SOPHIE \citep{perruchot2008SOPHIE} spectrographs, as well as photometric data from TESS Sectors 9, 36, 45, and 46 \citep{ricker2015tess}. The RV data from CORALIE and SOPHIE were sourced from \citet{smith2014wasp}.

We drew initial guesses for $P$, $T_{0}$, $\cos{i}$, $R_{\mathrm{p}}/R_{\star}$, $(R_{\star}+R_{\mathrm{p}})/a$, and $K$ (definitions given in Table \ref{tab:system_properties}) from the values derived in \citet{smith2014wasp}. All fitted parameters were allowed to vary and were initialized with uniform priors, as listed in Table \ref{tab:system_properties}. The two eccentricity parameters $\sqrt{e} \sin{\omega}$ and $\sqrt{e} \cos{\omega}$ were each initialized with a value of 0, and the two transformed quadratic limb-darkening coefficients $q_1$ and $q_2$ \footnote{The relations between the transformed ($q_1$, $q_2$) and physical ($u_1$, $u_2$) quadratic limb-darkening coefficients are defined by Equations 15 and 16 in \cite{Kipping2013eccq1q2}: $u_{1}=2 \sqrt{q_{1}} q_{2}$ and $u_{2}=\sqrt{q_{1}}\left(1-2 q_{2}\right)$.} for each of the TESS and NEID datasets (4 total coefficients) were initialized with a value of 0.5. We fit baseline RV offsets for the CORALIE, SOPHIE, and NEID datasets, allowing each to vary between $\pm20$ km/s. To support convergence, the three offsets were each initialized at 17.2 km/s after manually examining the data. The sky-projected spin-orbit angle $\lambda$ was initialized with a value of $0^{\circ}$ and allowed to vary between $\pm180^{\circ}$. The projected stellar rotational velocity $v\sin{i_*}$ was initialized with a value of $5.83$ km/s from our spectral fit and was allowed to vary between 0 and 20 km/s.

We ran an affine-invariant Markov Chain Monte Carlo (MCMC) analysis with 100 walkers to sample the posterior distributions of all model parameters. The best-fit model parameters and their associated 1$\sigma$ uncertainties, listed in Table \ref{tab:system_properties}, were extracted after obtaining 200,000 accepted steps per walker, where the first 10,000 steps were discarded as burn-in. Our results are in good agreement (that is, within $2\sigma$) with the associated values obtained by \citet{smith2014wasp}.

The best-fit joint model is shown in Figure \ref{fig:rv_joint_fit} together with each dataset included in the analysis, as well as the residuals of each fit. The fitted and derived parameters corresponding to this model are provided in Table \ref{tab:system_properties}. The WASP-106 system is consistent with a low spin-orbit angle, with $\lambda=6_{-16}^{+17}\,\degr$.

\begin{figure}
    \centering
    \includegraphics[width=1\linewidth]{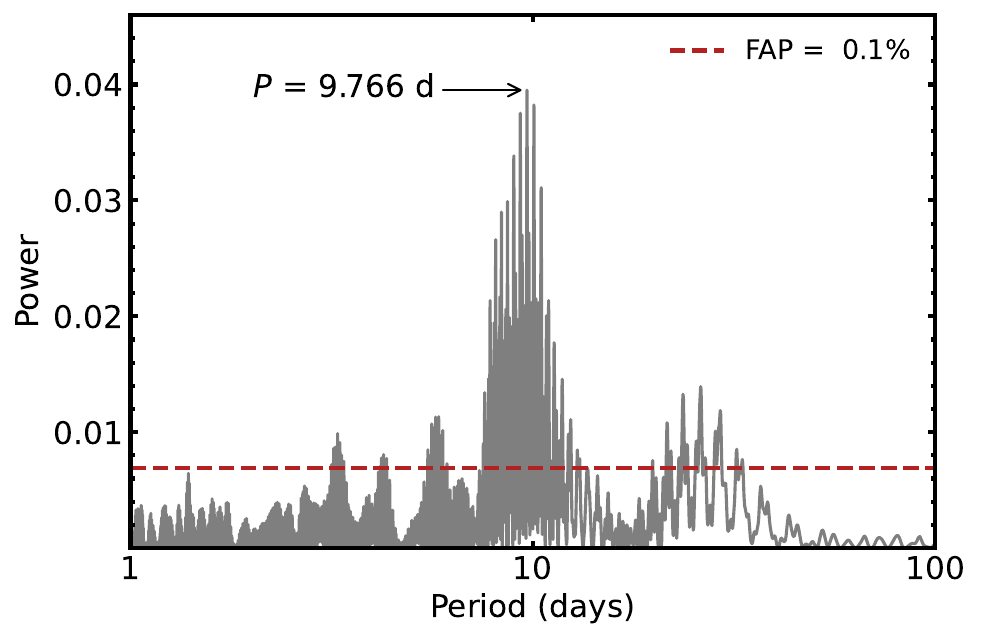}
    \caption{Lomb-Scargle periodogram of the TESS light curve data for WASP-106, with transit data masked out. A False Alarm Probability (FAP) level at 0.1\% is marked with a dashed red line. The highest-power peak corresponds to a period $P = 9.766$ days marked with an arrow, indicating the most likely stellar rotation period.}
    \label{fig:periodogram}
\end{figure}

Next, we leveraged TESS light curve data, in combination with our derived $\lambda$ constraint, to measure the 3D spin-orbit angle $\psi$. Our analysis incorporated two-minute cadence TESS light curve data from Sectors 9, 36, 45, and 46, spanning February 28th, 2019 to December 30th, 2021. Based on a Generalized Lomb-Scargle Periodogram (GLS, \citealt{Zechmeister2009}) analysis, we found $P_{\rm rot}=9.766\pm0.005$ days with a False Alarm Probability (FAP) of less than $0.1\%$, as shown in Figure\,\ref{fig:periodogram}. However, because latitudinal differential rotation enforces a lower limit of $10\%$ to the measurement precision  \citep{Epstein2014,Aigrain2015}, we ultimately adopted $P_{\rm rot}=9.77\pm0.98$ days. 

Combining this value with $v\sin {i_*}= 7.0_{-1.0}^{+1.1}$ km/s from our global fit, the stellar equatorial rotation velocity was derived as $v = \frac{2 \pi R_{*}}{P_{\rm rot}}=7.61\pm0.77$ km/s. The Bayesian inference method from \cite{Masuda2020stincl} and \cite{Hjorth2021} was then applied to $R_{*}$, $P_{\rm rot}$, and $\cos{i_*}$ to accommodate the interdependent parameters $v$ and $v\sin{i_*}$. The fitted parameters were \rstar, ${P_{\rm rot}}$, and $\cos i_{*}$, and uniform priors were applied to them. To achieve a conservative result, we adopted the suggested systematic uncertainties of $\sigma_{R_{*}} \approx 4.2\%$, which equates to 0.06 $R_{\odot}$. The final likelihood function is given as

\begin{equation}
\begin{aligned}
\mathcal{L} &= \left(\frac{R_{\star} / R_{\odot} - 1.47}{0.06}\right)^{2} + \left(\frac{P_{\mathrm{rot}} - 9.77 \mathrm{~d}}{0.98 \mathrm{~d}}\right)^{2} \\
&\phantom{=} + \left(\frac{v\sqrt{(1-\cos^{2}i_{*})} - 7.0 \mathrm{~km} / \mathrm{s}}{1.1 \mathrm{~km} / \mathrm{s}}\right)^{2},
\end{aligned}
\end{equation}
where $v = 2\pi\rstar/P_{\mathrm{rot}}$. Note that we adopt $\sigma_{R_*}=0.06R_{\odot}$ based on the systematic stellar parameter uncertainties suggested by \cite{Tayar2020}. 

We implemented the likelihood function using \texttt{PyMC3} \citep{pymc3} and ran MCMC sampling until the Gelman-Rubin statistic ($\hat R$) for each fitted parameter was less than 1.01.  We derived a stellar inclination posterior $\sin i_*=0.91\pm0.09$, or $i_*=89.72\pm25.15\degr$. Subsequently, the true stellar obliquity ($\psi$) was calculated using \citep{fabrycky2009exoplanetary}

\begin{equation}
    \cos \psi=\cos i_{*} \cos i+\sin i_{*} \sin i \cos \lambda,
\end{equation}
where $i_{*}$ is the stellar inclination and $i$ is the planet's orbital inclination. We ultimately obtained $\psi=26^{+12}_{-17}\degr$, which is consistent with near-alignment for WASP-106 b.



\begin{deluxetable*}{llllll}
\tablecaption{System properties derived for WASP-106.}
\label{tab:system_properties}
\tabletypesize{\scriptsize}
\tablehead{\colhead{Parameter} & \colhead{Description} & \colhead{Priors} & \colhead{Value} &  \colhead{$+1\sigma$} & \colhead{-1$\sigma$}}
\tablewidth{300pt}
\startdata
\vspace{-2mm}
& & & & & \\
Stellar Parameters$^{\dagger}$: & & & & & \\
~~~~$M_*$ \dotfill & Stellar mass ($M_\odot$) \dotfill &-&1.175  &0.082   &0.074 \\
~~~~$R_*$\dotfill &Stellar radius ($R_\odot$) \dotfill& -&  1.47& 0.016  &  0.017\\
~~~~$\log{g}$ \dotfill & Surface gravity (cm/s$^2$)  \dotfill&-&   4.49&  0.16 &  0.16\\
~~~~$[\rm{Fe/H}]$ \dotfill & Metallicity (dex) \dotfill&- &  -0.02 &  0.10 & 0.10\\
~~~~$T_{\rm eff}$ \dotfill & Effective temperature (K) \dotfill&-& 6002 & 164 & 164 \\
\vspace{2mm}
~~~~$v\sin i_*$\dotfill & Projected stellar rotational velocity (km/s)\dotfill & $\mathcal{U}(5.83;0.0;20.0)$ & 7.0 & 1.1 & 1.0 \\
Planetary Parameters: & & & & & \\
~~~~$R_{\mathrm{p}} / R_\star$\dotfill & Planet-to-star radius ratio\dotfill & $\mathcal{U}(0.0782376;0;1)$\tablenotemark{$*$} & $0.07559$ & 0.00072 & 0.00087 \\
~~~~$(R_\star + R_\mathrm{{p}}) / a$\dotfill & Sum of radii divided by the orbital semimajor axis\dotfill & $\mathcal{U}(0.0762043;0;1)$ & $0.0816$ & 0.0024 & 0.0021 \\
~~~~$\cos{i}$\dotfill & Cosine of the orbital inclination\dotfill & $\mathcal{U}(0.0089;0;1)$ & 0.0297 & 0.0063 & 0.0103 \\
~~~~$T_{0}$\dotfill & Mid-transit epoch ($\mathrm{BJD}-2457000$)\dotfill & $\mathcal{U}(977;644;998)$ & 977.9796 & 0.0014 & 0.0014 \\
~~~~$P$\dotfill & Orbital period (days)\dotfill &$\mathcal{U}(9.289715;8.28;10.28)$ & 9.289699 & 1e-05 & 1e-05 \\
~~~~$K$\dotfill & Radial velocity semi-amplitude ($\mathrm{m/s}$)\dotfill & $\mathcal{U}(165.3;0;1000)$ & 164.7 & 4.4 & 4.4 \\
~~~~$\sqrt{e} \cos{\omega}$\dotfill & Eccentricity parameter 1\dotfill & $\mathcal{U}(0;-1.0;1.0)$ & -0.063 & 0.084 & 0.067 \\
~~~~$\sqrt{e} \sin{\omega}$\dotfill & Eccentricity parameter 2\dotfill & $\mathcal{U}(0;-1.0;1.0)$ & 0.091 & 0.134 & 0.137 \\
~~~~$q_{\rm1, TESS}$\dotfill & Quadratic limb darkening coefficient 1, TESS\dotfill & $\mathcal{U}(0.5;0.0;1.0)$ & 0.095 & 0.092 & 0.041 \\
~~~~$q_{\rm2, TESS}$\dotfill & Quadratic limb darkening coefficient 2, TESS\dotfill & $\mathcal{U}(0.5;0.0;1.0)$ & 0.50 & 0.33 & 0.31 \\
~~~~$q_{\rm1, NEID}$\dotfill & Quadratic limb darkening coefficient 1, NEID\dotfill & $\mathcal{U}(0.5;0.0;1.0)$ & 0.49 & 0.31 & 0.29 \\
~~~~$q_{\rm 2, NEID}$\dotfill & Quadratic limb darkening coefficient 2, NEID\dotfill & $\mathcal{U}(0.5;0.0;1.0)$ & 0.56 & 0.30 & 0.36 \\
~~~~$\Delta_{\rm RV, CORALIE}$\dotfill & RV offset, CORALIE (km/s)\dotfill & $\mathcal{U}(17.2;-20.0;20.0)$ & 17.248 & 0.004 & 0.004 \\
~~~~$\Delta_{\rm RV, SOPHIE}$\dotfill & RV offset, SOPHIE (km/s)\dotfill & $\mathcal{U}(17.2;-20.0;20.0)$ & 17.189 & 0.006 & 0.006 \\
~~~~$\Delta_{\rm RV, NEID}$\dotfill & RV offset, NEID (km/s)\dotfill & $\mathcal{U}(17.2;-20.0;20.0)$ & 17.189 & 0.004 & 0.004 \\
~~~~$\lambda$\dotfill & Sky-projected spin-orbit angle ($\degr$)\dotfill &  $\mathcal{U}(0;-180.0;180.0)$ & 6 & 17 & 16 \\
\vspace{2mm}
Derived Parameters: & & & & & \\
~~~~$R_\mathrm{p}$ \dotfill & Planetary radius ($\mathrm{R_{J}}$)\dotfill &- & 1.080 & 0.016 & 0.017 \\
~~~~$M_\mathrm{p}$\dotfill & Planetary mass  ($\mathrm{M_{J}}$)\dotfill &- & 1.93 & 0.15 & 0.15 \\
~~~~$b$\dotfill & Impact parameter\dotfill &- & 0.387 & 0.074 & 0.136 \\
~~~~$T_\mathrm{14}$\dotfill & Transit duration (h)\dotfill &- & 5.334 & 0.040 & 0.038 \\
~~~~$\delta$\dotfill & Transit depth\dotfill &- & 6.204 & 0.086 & 0.071 \\
~~~~$a$\dotfill & Semimajor axis (au)\dotfill &- & 0.0901 & 0.0026 & 0.0027 \\
~~~~$i$\dotfill & Inclination  ($\degr$)\dotfill &- & 88.30 & 0.59 & 0.36 \\
~~~~$e$\dotfill & Eccentricity\dotfill &- & 0.023 & 0.027 & 0.016 \\
~~~~$\omega$\dotfill & Argument of periastron ($\degr$)\dotfill &- & 128 & 93 & 37 \\
~~~~$u_\mathrm{1, TESS}$\dotfill & Limb darkening parameter 1, TESS\dotfill &- & 0.30 & 0.12 & 0.15 \\
~~~~$u_\mathrm{2, TESS}$\dotfill & Limb darkening parameter 2, TESS\dotfill &- & 0.00 & 0.25 & 0.16 \\
~~~~$u_\mathrm{1, NEID}$\dotfill & Limb darkening parameter 1, NEID\dotfill &- & 0.69 & 0.53 & 0.46 \\
~~~~$u_\mathrm{2, NEID}$\dotfill & Limb darkening parameter 2, NEID\dotfill &- & -0.07 & 0.45 & 0.40 \\
\enddata
\tablenotetext{\dagger}{The resulting uncertainties of stellar parameters did not account for systematic errors \citep{Tayar2020}.}
\tablenotetext{$*$}{$\mathcal{U}(x;a;b)$ is a uniform prior with initial guess $x$ and lower and upper limits $a$ and $b$, respectively.}


\end{deluxetable*}


\section{Tidal Realignment Timescales}
\label{section:tidal_realignment_timescales}

Next, we verified the expected tidal realignment timescales for WASP-106 b to demonstrate whether the system could have been realigned from a misaligned state within its lifetime. For cooler stars (below the Kraft break) with significant convective envelopes, the convective tidal realignment timescale $\tau_{\rm CE}$ is given by \citep{zahn1977tidal, albrecht2012obliquities}
\begin{equation}
    \frac{1}{\tau_{\rm CE}} = \frac{1}{10 \times 10^9 \, \rm{yr}} \bigg(\frac{M_{\rm p}}{M_{*}}\bigg)^2 \bigg(\frac{a/R_*}{40}\bigg)^{-6}.
\end{equation}
For hotter stars (above the Kraft break) with much less appreciable convective envelopes, the radiative realignment timescale $\tau_{\rm RA}$ is given by \citep{zahn1977tidal, albrecht2012obliquities}
\begin{multline}
    \frac{1}{\tau_{\rm RA}} = \frac{1}{1.25 \times 5 \times 10^9 \, \rm{yr}} \bigg(\frac{M_{\rm p}}{M_*}\bigg)^2 \\
    \times \bigg(1 + \frac{M_{\rm p}}{M_*}\bigg)^{5/6} \bigg(\frac{a/R_*}{6}\bigg)^{-17/2}.
\end{multline}

 These equations have been empirically calibrated using observations of binary star systems \citep{zahn1977tidal}. This means that their application in this context includes an implicit assumption that planet-star systems follow a similar tidal realignment mechanism to that of binary star systems. Therefore, we warn that their use is not intended to provide a precise value for the expected tidal realignment timescale, but, rather, an order-of-magnitude estimate. 

Given our measured stellar metallicity $\mathrm{[Fe/H]}=-0.02\pm0.10$ (Table \ref{tab:system_properties}), we anticipate that the temperature of WASP-106 likely falls just below the metallicity-dependent Kraft break, which is expected to lie in the range $6100 \mathrm{K} \leq T_{\rm eff} \leq 6200$ K based on Figure 9 in \citet{Spalding_2022}. Nevertheless, because WASP-106 lies directly along the border delineating Kraft break, with $T_{\rm eff} = 6002 \pm 164 \,\rm K$, we compute both $\tau_{\rm CE}$ and $\tau_{\rm RA}$ for thoroughness. 

Using the values from Table \ref{tab:system_properties}, we find that $\tau_{\rm CE} = 5.26 \times 10^{12} \, \rm{yr}$ and $\tau_{\rm RA} = 2.06 \times 10^{18} \, \rm{yr}$. This result suggests that the WASP-106 system was likely never significantly misaligned, as realigning WASP-106 b, regardless of whether the host star had a significant convective envelope or not, would have taken several orders of magnitude more years than the age of the Universe. Our result strengthens the hypothesis that warm Jupiters commonly form in aligned configurations \citep{rice2022tendency} even in systems along the Kraft break, suggesting that warm Jupiters may generally form more quiescently than hot Jupiters.

\section{Discussion}
\label{section:discussion}

On a broad scale, the SOLES survey aims to delineate the origins of spin-orbit misalignments by examining the spin-orbit distribution of wide-orbiting exoplanets. Because most spin-orbit observations to date -- including the one presented in this work for WASP-106 b -- have been made for transiting giant planets, we focus on hot and warm Jupiters within this section. 

Figure \ref{fig:hot_warm_comparison} shows a comparison of the sky-projected spin-orbit distribution for hot (top panel) and warm (bottom panel) Jupiters, distinguishing between planets as a function of stellar $T_{\rm eff}$ and multiplicity of the host star system.\footnote{This figure includes all systems with $\lambda$ measurements in the TEPcat catalogue \citep{southworth2011homogeneous} as of 7/20/2023.} Systems with one or more stellar companions were identified by (1) searching for systems with $\texttt{sy\_snum}>1$ in the NASA Exoplanet Archive and (2) applying the criteria outlined in \citet{El_Badry_2021} to check for any bound companions within $10\arcmin$ of the primary resolved by the \textit{Gaia} DR3 catalogue. We found no stellar companions bound to the WASP-106 system.

As established in previous studies \citep{winn2010hot, schlaufman2010evidence}, we recover the known trend that hot Jupiters around hot stars are misaligned at a relatively high rate, whereas those around cool stars are typically aligned. We also find that, with the most updated sample, all warm Jupiters in single-star systems with spin-orbit measurements to date remain at or near spin-orbit alignment, as initially found in \citet{rice2022tendency}. WASP-106 b is consistent with this pattern.

The host star's position along the lower edge of the Kraft break makes this measurement particularly interesting: the system's alignment, in combination with other aligned warm Jupiter systems around the Kraft break, may suggest the absence of a stellar obliquity transition between hot and cool stars hosting warm Jupiters. However, further observations in this crucial parameter space will be necessary to definitively demonstrate the presence or absence of this transition. We suggest three potential scenarios that are each consistent with the most updated stellar obliquity distribution shown in Figure \ref{fig:hot_warm_comparison}:



\begin{figure*}
    \centering
    \includegraphics[width=.9\linewidth] {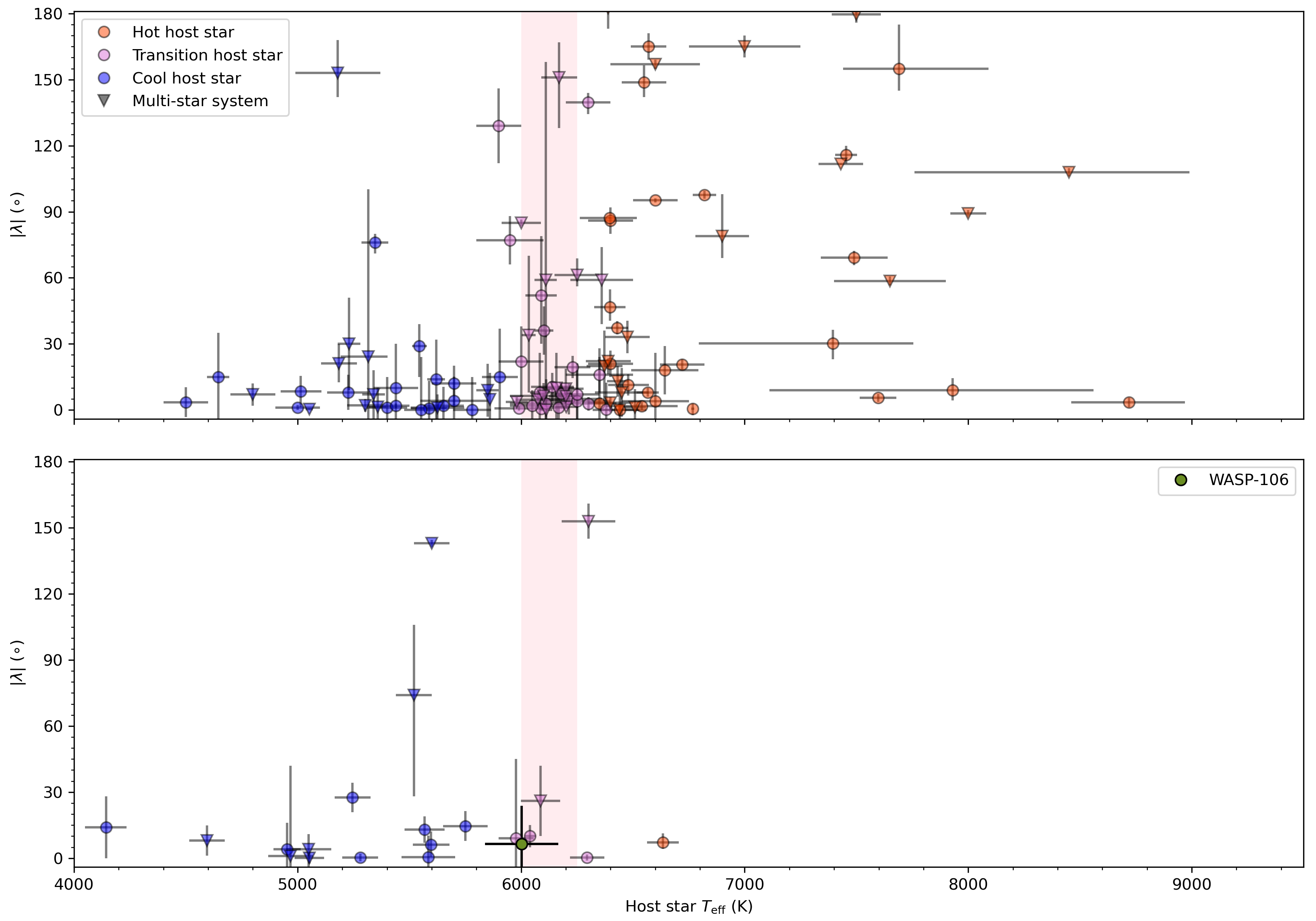}
    \caption{Stellar obliquity distribution for hot Jupiters (top) and warm Jupiters (bottom). The range used for the Kraft break, marked in pink, is $6000 \leq T_{\rm eff} \leq 6250\,\rm K$, to account for the potential variation of the Kraft break among individual host stars \citep{Spalding_2022}. We define a hot Jupiter as any planet with $a/R_{\star} < 11$ and $M \geq 0.4 \,\rm M_J$, and a warm Jupiter as any planet in the same mass range with $a/R_{\star} \geq 11$. Single-star systems are marked with a circle, and multi-star systems with a triangle. Values were obtained from the TEPCat catalog \citep{southworth2011homogeneous} on 7/20/23.}
    \label{fig:hot_warm_comparison}
\end{figure*}


\begin{enumerate}
    
\item Hot and warm Jupiters form through distinct channels, with warm Jupiters forming quiescently and being initially aligned, while hot Jupiters form violently and are initially misaligned. Only hot Jupiters orbiting cool stars would be tidally realigned, producing the current stellar obliquity distribution for hot Jupiters \citep[e.g.][]{albrecht2012obliquities, rice2022origins}. 

In this scenario, we would expect to observe relatively low stellar obliquities for warm Jupiters around both cool and hot stars. 

\item Both hot and warm Jupiters orbiting cool stars have undergone quiescent formation histories and are therefore initially aligned, while those orbiting hot stars have experienced more violent formation processes and are initially misaligned. Because more massive, hotter stars tend to have a higher rate of stellar multiplicity \citep{duchene2013_stellar_multi, Yang2020_populations}, they are more likely to encounter the perturbing influence of a stellar companion, which can produce a primordial disk misalignment  \citep{batygin2012primordial}. These hotter stars are also more likely to host more massive protoplanetary disks \citep{andrews2013mass}, which may be more likely to form multiple Jupiters capable of inducing misalignment through post-disk dynamical sculpting \citep{WuD2023}. 

In this case, the current spin-orbit distribution would directly reflect the planet formation process, with tides playing a lesser role in altering stellar obliquities over time \citep{Hixenbaugh2023}. Accordingly, we would expect that the population of warm Jupiters orbiting hot stars would follow a comparable spin-orbit distribution to that of the hot Jupiters orbiting hot stars. Because there are only a few warm Jupiters orbiting hot stars with measured obliquities, the currently observed alignment in these systems may simply reflect small-number statistics.

\item Hot Jupiters orbiting cool stars form and evolve in a similar, quiescent manner to warm Jupiters, and they are therefore initially aligned \citep{WuD2023, Hixenbaugh2023}. Only hot Jupiters orbiting hot stars have undergone violent formation histories, resulting in initial misalignment. Hot Jupiters around hot stars would then represent a subset of planets that initially formed as longer-period Jupiters but that were dynamically excited and that tidally circularized within the system lifetime (as in e.g. the framework proposed in \citet{WuD2023}).

In this case, as in Scenario 1, we would expect to observe relatively small spin-orbit angles for warm Jupiters around both hot and cool stars.
\end{enumerate}

To distinguish between these three scenarios, it is necessary to expand the sample of warm Jupiter spin-orbit angles to include more measurements in systems with host stars along and above the Kraft break \citep[e.g.][]{sedaghati2023_TOI677}. The presented measurement of WASP-106 b supports this goal and builds toward future population studies that will demonstrate how warm Jupiters fit into a broader context. Ultimately, the stellar obliquity distribution for warm Jupiter systems holds great promise to provide insights into whether the current geometries of hot and warm Jupiter systems are primarily an outcome of formation processes, or whether misalignments are instead a consequence of subsequent dynamical evolution.


\section{Acknowledgements}
\label{section:acknowledgements}
M.R. and S.W. thank the Heising-Simons Foundation for their generous support. M.R. acknowledges support from Heising-Simons Foundation Grant \#2023-4478. S.W. acknowledges support from Heising-Simons Foundation Grant \#2023-4050. This research was supported in part by Lilly Endowment, Inc., through its support for the Indiana University Pervasive Technology Institute. 

\vspace{5mm}
While this paper was in the review process, it was brought to our attention that another team conducted a separate Rossiter-McLaughlin measurement of WASP-106 b \citep{harre2023orbit}. Our analysis was conducted fully independently of this result, and both spin-orbit measurements are consistent with each other.

\software{\texttt{numpy} \citep{oliphant2006guide, walt2011numpy, harris2020array}, \texttt{matplotlib} \citep{hunter2007matplotlib}, \texttt{pandas} \citep{mckinney2010data}, \texttt{scipy} \citep{virtanen2020scipy}, \texttt{allesfitter} \citep{gunther2020allesfitter}, \texttt{emcee} \citep{foremanmackey2013}, \texttt{iSpec} \citep{blanco2014ispec, blanco2019ispec}, \texttt{EXOFASTv2} \citep{eastman2019exofast}}

\facility{NEID/WIYN, TESS, SOPHIE, CORALIE, NASA Exoplanet Archive, Extrasolar Planets Encyclopaedia}

\facility{This work has made use of data from the European Space Agency (ESA) mission
{\it Gaia} (\url{https://www.cosmos.esa.int/gaia}), processed by the {\it Gaia}
Data Processing and Analysis Consortium (DPAC,
\url{https://www.cosmos.esa.int/web/gaia/dpac/consortium}). Funding for the DPAC
has been provided by national institutions, in particular the institutions
participating in the {\it Gaia} Multilateral Agreement.}

\bibliography{bibliography}
\bibliographystyle{aasjournal}

\end{document}